# Growth of metal nanoparticles in hydrocarbon atmosphere of arc discharge

S. Musikhin[*], V. Nemchinsky, and Y. Raitses

Princeton Plasma Physics Laboratory, Princeton, NJ 08543, United States of America

## 1 Abstract

A direct current (DC) arc discharge is a widely used method for large-scale production of metal nanoparticles, core-shell particles, and carbon nanotubes. Here, we explore the growth of iron nanoparticles in a modified DC arc discharge. Iron particles are produced by the evaporation of an anode, made from low-carbon steel. Methane admixture into argon gas serves as a carbon source. Electron microscopy and elemental analysis suggest that hydrocarbons decompose on iron clusters forming a carbon shell, which inhibits iron particle growth until its full encapsulation, at which point the iron core growth is ceased. Experimental observations are explained using an aerosol growth model. The results demonstrate the path to manipulate metal particle size in a hydrocarbon arc environment.

## 2 Introduction

Metal nanoparticles possess attractive magnetic, optical, and catalytic properties, which depend on the particle size [1–3]. This motivates developing tools to control the size distribution of metal particles during their synthesis. Creating a carbon shell over the metal core is one of these tools. For example, chemical vapor deposition (CVD) studies demonstrated that once the carbon shell is formed the metal core growth is ceased [4,5].

Historically, however, the motivation for encapsulating metal nanoparticles was rather to protect the metal core from oxidation and agglomeration and to provide chemical and thermal stability, so that particles retain their properties in different environments [6]. Applications of such core-shell particles vary from magnetic fluids [7], syngas conversion [8], hydrogen evolution reaction [9], and drug delivery [10], to carbon nanotubes (CNT) synthesis [11].

*Corresponding author
smusikhin@pppl.gov

One of the dominant methods for single-step production of metal-carbon nanoparticles at a large scale is the direct current (DC) arc discharge [12–16], although other techniques have been used [17–22]. In the DC arc, the discharge is maintained between two electrodes, where an anode is usually made of graphite, which ablates and provides a feedstock of carbon for nanomaterials synthesis. Metal is typically added as a powder of microparticles, which evaporate in the hot arc and nucleate as nanoparticles.

Understanding and controlling metal particle growth is also relevant for arc discharge synthesis of CNT [23–28]. For example, catalyst particles larger than 5 nm are associated with the formation of multi-walled carbon nanotubes (MWCNT), while single-walled carbon nanotubes (SWCNT) typically require catalysts under 3—5 nm [29–32]. Although carbon coverage is commonly assumed to poison the catalyst, alternative experimental observations exist. Nagatsu et al. [11], using a DC arc discharge, prepared Ni and Fe particles encapsulated in several graphene layers, which they later used as catalysts to grow MWCNT in a CVD reactor. Schunemann et al. [5], observed growing tubes while they were surrounded by an amorphous carbon shell, and concluded that carbon feedstock can permeate through the shell and reach catalyst particles.

In this work, we utilize a modified arc approach where carbon feedstock is supplied by the decomposition of hydrocarbon gas ($CH_4$) in the hot arc core, while the evaporating steel anode provides the metal (Fe). The use of a gaseous hydrocarbon instead of a consumable graphite electrode makes this method continuous and scalable for industrial production, which was successfully commercialized for carbon black [33] and CNTs [34]. Importantly, methane is the second most abundant anthropogenic greenhouse gas that is also 28 times as potent as $CO_2$ at trapping heat in the atmosphere. By utilizing methane to produce high-value solid materials and hydrogen (not investigated here), we aim to make the process more sustainable. Given the scalability and sustainability of methane arc discharge to obtain high-value nanomaterials, it is important to develop an understanding of how metal aerosols grow in such environments.

The synthesis of metal-carbon core-shell particles using methane arc discharge was realized before [35–38]. Some studies also explored how the methane atmosphere affects particle size. Dong et al. compared the sizes of particles synthesized by evaporating an iron anode at different $CH_4$ pressures [35]. The authors observed that the mean particle size decreased as methane pressure increased from 13.3 to 40 kPa. While the effect was clearly observed, its interpretation is not straightforward. First, there was no direct comparison of particles synthesized with and without methane, while keeping other conditions constant. Second, the authors compared the overall sizes of particles rather than the iron cores, leaving an option that changes were due to the carbon shell thickness. Finally, experiments were conducted at different pressures, which could have impacted plasma characteristics and aerosol growth kinetics [39,40]. Hao et al. compared the sizes of copper particles encapsulated with carbon that were synthesized using He/$H_2$ (1:1) and He/$CH_4$ (1:1) gas mixtures [36]. They found that in the presence of $CH_4$, carbon limited the copper particles aggregation and hence, reduced their size. However, when replacing a large part of the gas mixture with another gas (e.g., half of the mixture from $H_2$ to $CH_4$), multiple arc parameters are expected to be affected due to different gas properties, such as specific heat capacity, and thermal and electrical conductivity [41]. Furthermore, the inconsistent particle sizes reported throughout the study complicate the interpretation of the results.

Here, we unambiguously demonstrate in an experiment and explain with an aerosol growth model that a carbon coating indeed inhibits the growth of iron nanoparticles formed from iron vapor in the Ar/$CH_4$ arc and that this effect alone is sufficient to obtain ultra-fine metal nanoparticles of only a couple of nanometers in size. We note that implementation of other size control knobs, such as adjusting the gas flowrates to vary species concentration and residence time or quenching to terminate the reaction kinetics, is limited in arcs, because of instabilities [42,43] and strong gradients of species densities, temperature, and pressure [44].

## 3 Experimental setup and procedures

Experiments were carried out in a stainless-steel vacuum chamber (Fig. 1 a). Electrodes, shown in Fig. 1 b, were oriented vertically inside the chamber, with a 6.4 mm diameter cathode on top (2% ceriated tungsten) and a 9.5 mm diameter anode underneath (A36 steel, Fe > 99 wt.%, C 0.06 wt.%). The anode was placed on a positioning stage so that the interelectrode gap was adjusted by a stepper motor. We aimed to keep a constant gap of 2—3 mm: electrodes separated enough, so that the melting anode does not weld to the cathode, but not too far from each other to stabilize the discharge. Furthermore, the constant interelectrode gap among multiple experiments suggests similar temperature fields and electrode ablation rates [44], and thus, allows for particle size comparison. The gap was continuously monitored with a camera and adjusted if necessary.

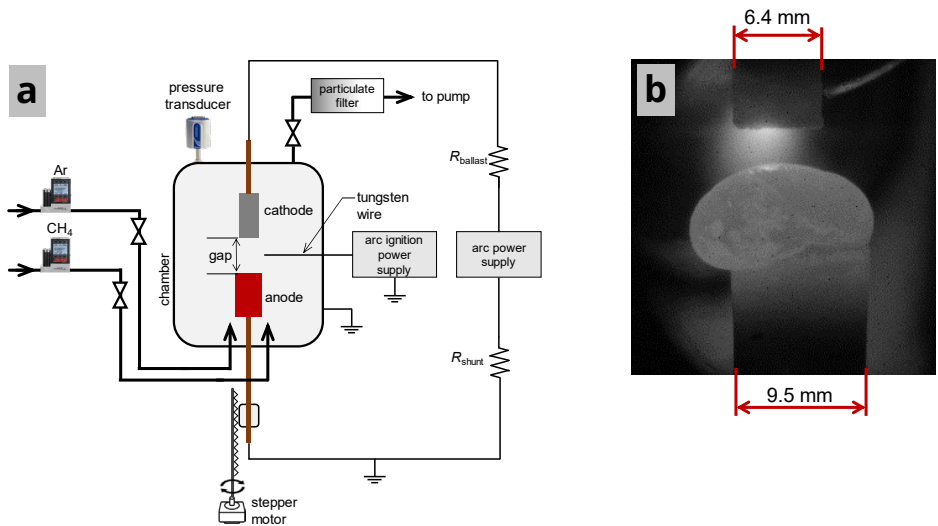

Fig. 1. a) Diagram of the experimental setup; b) Image of an arc in operation showing electrode diameters and the glowing spot size. The tungsten cathode (top in b) is fixed while the steel anode (bottom in b) position is adjusted with a stepper motor.

The reactor chamber was first pumped down to ~1 Pa (~10 mTorr) and subsequently filled with a working gas mixture to a pressure of 67 kPa (500 Torr). We investigated two cases: Ar (5.0 purity) flow of $916 \pm 5$ sccm; and a mixture of Ar ($916 \pm 5$ sccm) and $CH_4$ ($22.9 \pm 0.1$ sccm, 4.0 purity) flows, which corresponds to a 2.4 wt.% $CH_4$ in Ar. All gas flows were controlled by calibrated mass flow controllers (Alicat Scientific). The pressure inside the chamber was maintained at $67 \pm 4$ kPa ($500 \pm 30$ Torr) and monitored with a pressure transducer (MKS, Baratron 221A). The reactor was checked for leaks using a helium leak detector (Pfeiffer Vacuum), a typical leak was measured to be $10^{-7}..10^{-8}$ Torr l s$^{-1}$. The arc was ignited by a spark from a tungsten wire biased to 2 kV using a Bertan Associates Inc. (205A-05R) high-voltage power supply. The discharge was sustained with a Sorenson SGA100X100C-1AAA 100 V/100 A power supply, operated at a constant current of 33 A. The voltage measured across the two electrodes was 10—12 V in the case of pure Ar, and 12—16 V in the case of Ar/CH$_4$ mixture. The arc current was measured across a 2 mΩ shunt resistor, denoted in Fig. 1 as $R_{shunt}$. The current value was chosen based on the previous research in similar arcs (although with graphite electrodes), which showed that the low ablation regime at smaller currents (below 55 A for graphite) is more stable and allows for longer arc operation [45]. A typical run lasted for 1.5 hours, limited by chamber heating. After turning the plasma off, the reactor was evacuated with the pump, filled with Ar to 80 kPa (600 Torr), and left for at least an hour to cool down.

Particle samples were collected for *ex situ* analysis in two ways, using i) a PTFE-membrane particulate filter (Cobetter, MFPT-2247) installed stationary in the exhaust vacuum line; ii) a TEM grid (200-mesh Cu with a lacey carbon film) placed on the chamber bottom. In the former case, the filter collected particles carried with the gas flow to a vacuum pump. In the latter case, particles are deposited on the grid under gravity and free convection flow.

Particle morphology, size distribution, and chemical composition were analyzed using a scanning transmission electron microscope (STEM, FEI Talos FX200, 200 kV) with integrated energy dispersive spectroscopy (EDS, 20 kV). Powder collected on a particulate filter was scraped off, dispersed in absolute ethanol (≥99.5% purity), and ultrasonicated for 10 min to form a nanocolloid. One droplet of the nanocolloid was then deposited on a TEM grid for analysis. We observed no difference between such samples and the ones collected directly on a TEM grid placed inside the reactor.

## 4 Experimental results

Fig. 2 shows typical TEM images of nanoparticles synthesized in the arc reactor either in pure Ar (Fig. 2 a—d) or in an Ar/CH$_4$ mixture (Fig. 2 e—h). In both cases, fractal aggregates consisting of primary spherical particles were formed, which is typical for gas-phase syntheses [46]. In pure Ar atmosphere, primary particles are mainly composed of Fe and O, as shown in Fig. 3 top row by high-angle annular dark-field (HAADF) STEM and energy dispersive x-ray spectroscopy (EDS). Considering a small but non-zero leak in the chamber (a typical leak rate of $10^{-7}..10^{-8}$ Torr l s$^{-1}$) and that iron nanoparticles are prone to oxidation even at room temperature [47], particles could have oxidized inside of the reactor as well as outside, when extracted for analysis.

Fig. 2 e—h shows dark aggregates enveloped in a light amorphous material. The chemical composition (Fig. 3 bottom row) suggests that the darker particles consist of iron while the shell is an amorphous carbon. EDS maps and TEM images show that, when adding methane to the system, iron particles get entirely coated with carbon. In some images (see Supplementary Information), discrete, non-aggregated iron particles were found, which were still embedded in amorphous carbon. No such discrete iron particles were observed in the case of pure Ar, only fractal aggregates. It suggests that carbon shells have formed early enough to prevent at least some iron clusters from coagulating and forming aggregates. That could have also limited iron particle growth by iron atoms adsorption and clusters coalescence. To check this, particle size distributions were experimentally measured for both working mixtures, with and without methane.

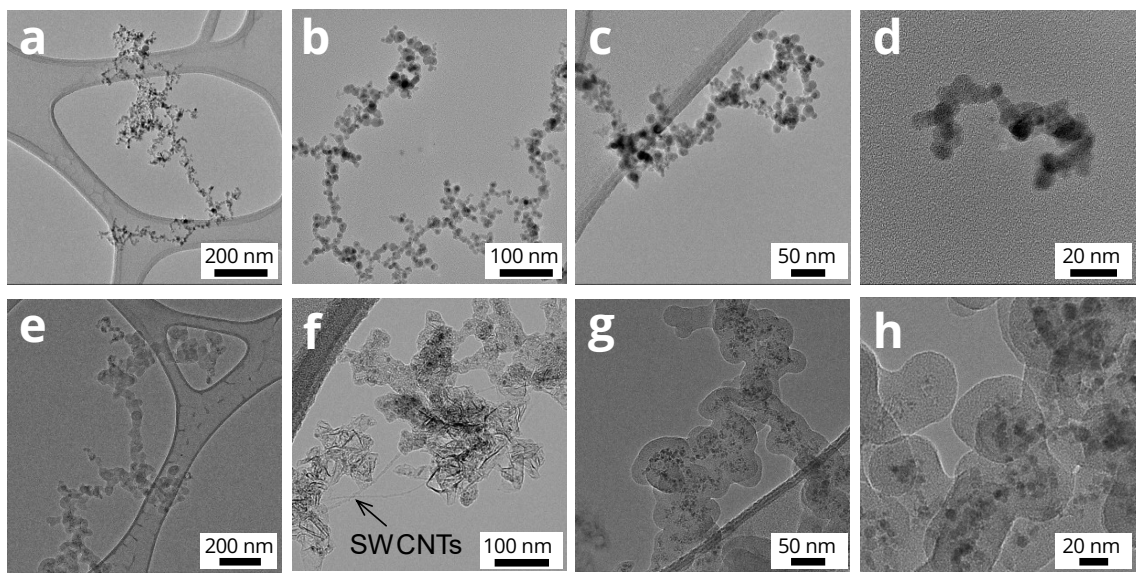

Fig. 2. Typical TEM images of particles synthesized in pure Ar (a—d) or a mixture of 2.4 wt. % $CH_4$ in Ar (e—h). Methane addition led to iron-carbon core-shell particles formation.

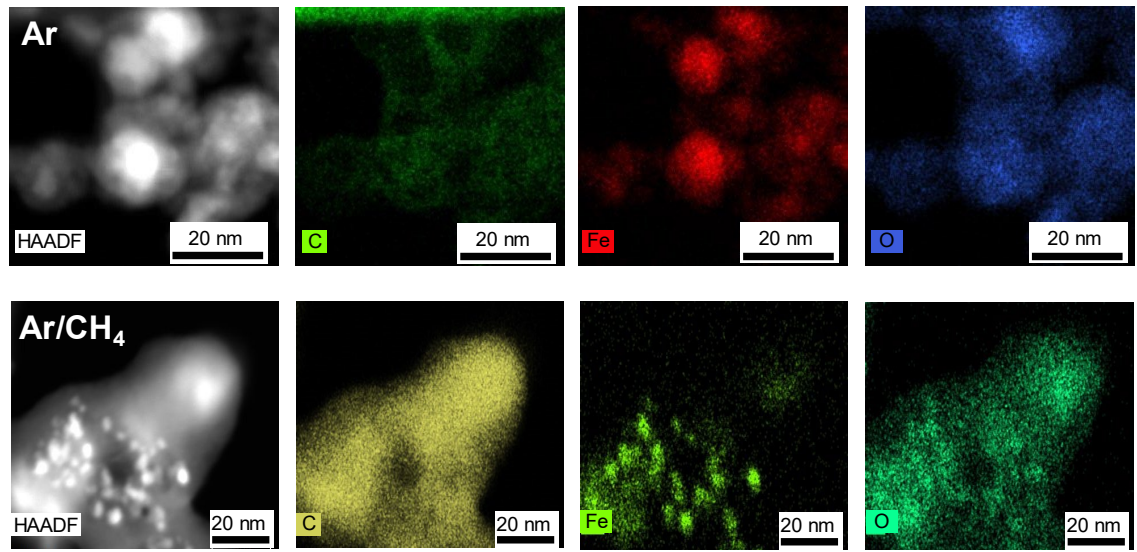

Fig. 3. HAADF STEM image and EDS elemental maps of aggregates synthesized in pure Ar (top row) or a mixture of 2.4 wt. % $CH_4$ in Ar. When adding methane, iron particles got covered by carbon.

Iron particle size distributions are plotted in Fig. 4 as histograms of the probability density functions (PDF) along with lognormal fits, which yield count median diameters (CMD) and geometric standard deviations (GSD). More than 150 particles were measured to build each histogram. Multiple samples were analyzed but only two extreme cases, with the smallest and largest CMDs for each gas mixture, are shown in Fig. 4. Iron core size decreased from 8—11 nm in pure Ar to 3—5 nm when adding 2.4 wt. % $CH_4$ to Ar. This experimentally confirms that the carbon shell formation limits iron core growth. Furthermore, in the case of Ar/$CH_4$ mixture, obtained catalyst particles were under 5 nm, which is expected to favor SWCNT formation relative to MWCNT and soot [29–31,48]. However, the efficient synthesis of SWCNT was not the goal of this work, hence, we did not manipulate any other process parameters, e.g., pressure, gas temperature, precursor/catalyst types, and their concentrations [49]. This explains why only a small number of SWCNT was detected in the Ar/$CH_4$ sample (Fig. 2 f). The efforts to improve SWCNT yield in the methane arc discharge are currently ongoing.

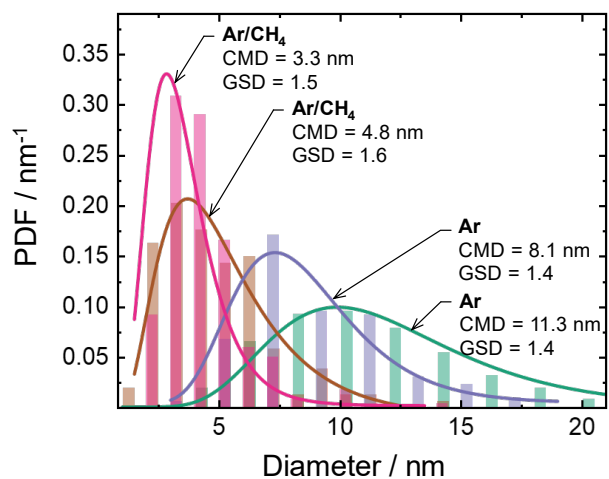

Fig. 4. Iron particle size probability density functions (PDF) and corresponding lognormal fits. CMD: count median diameter, GSD: geometric standard deviation. For Ar/CH$_4$ mixtures, the methane fraction was 2.4 wt. %. The addition of methane limited iron particle growth: from 8—11 nm in Ar to 3—5 nm in Ar/CH$_4$ mixture.

## 5 Discussion

To explain experimental observations, we refer to the aerosol growth model from Ref. [50], which uses a commonly accepted nucleation model by Girshick and Chiu [51], and accounts for condensation, evaporation, and coalescence. We extend the model with the possibility for hydrocarbon decomposition on the metal cluster surface, which leads to carbon diffusion in particle and shell formation. For the exact formulas and notations, we refer a reader to the original model publication [50], while here, for the sake of the "proof of principle" analysis, simplified phenomenological terms are used. Schematics of the considered processes are depicted in Fig. 5.

To follow the evolution of clusters size in time, we introduce the mean cluster volume defined as the ratio of the total clusters volume density, $N_{volume}$, to the cluster density, $N_{cluster}$. This monodisperse size approximation is often used in cluster growth models, e.g., in Ref. [50]. The rate change of $N_{volume}$, is given by

$$\frac{\partial N_{\text{volume}}}{\partial t} = V_{\text{nucleation}} + V_{\text{condensation}}(1 - \theta), \quad (1)$$

where the first term on the RHS represents volume growth due to clusters nucleation, while the second due to condensation of atoms on the cluster. A parameter $\theta$ is the fraction of the cluster surface occupied by the adsorbed carbon atoms, e.g., $\theta = 0$ corresponds to a clean metal surface, and $\theta = 1$ corresponds to a surface fully encapsulated in carbon. It is assumed that once the adsorption site is occupied by carbon, it is not available for iron atoms. Hence, $1-\theta$ is the part of the cluster surface that is still accessible for iron atoms adsorption. Note that Eq. 1 does not include coalescence as it does not change the total clusters volume.

The rate change of cluster density is given by

$$\frac{\partial N_{\text{cluster}}}{\partial t} = N_{\text{nucleation}} - N_{\text{coalescence}}(1 - \theta)^2. \quad (2)$$

Cluster density increases due to nucleation of new clusters and decreases due to coalescence. Parameter $(1-\theta)^2$ describes a probability of two iron clusters coalescing, where each cluster has a $1-\theta$ part of its surface not occupied by carbon.

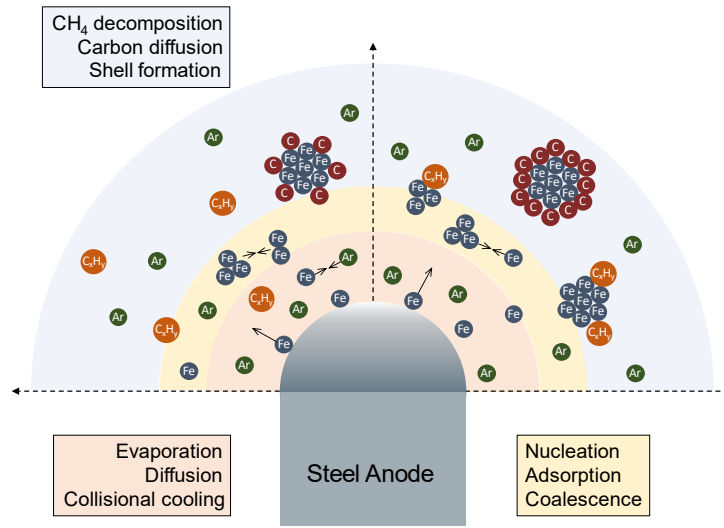

Fig. 5. Schematics of iron particles growth near evaporating steel anode in a hydrocarbon atmosphere. Not up to scale. Considered processes: i) Evaporation of iron from a hot molten anode surface; ii) Iron vapor diffusion into the cold argon gas, iron vapor cooling, and nucleation of clusters; iii) Clusters growth by iron atoms condensation and coalescence; iv) Hydrocarbons decomposition in the hot arc; v) Carbon adsorption, diffusion, and shell formation.

First, let us briefly describe a process of iron aerosol growth by inert condensation. As iron atoms evaporate from the anode surface, they diffuse into and collide with argon buffer gas and cool down. At the same time, when moving away from the anode, the buffer gas temperature drops leading to iron atoms supersaturation, which prompts clusters nucleation [51]. Clusters grow by iron atoms adsorption and coalescence. Following our notation, both processes occur with $\theta = 0$ in Eqs. 1 and 2, due to the absence of hydrocarbons. Adsorption consumes iron atoms, decreasing the level of supersaturation. Furthermore, atoms and clusters diffuse into space, which reduces their densities as $1/R^3$, where $R$ is the distance from the evaporation surface. At some distance, iron atom density drops to a level below supersaturation, so nucleation stops. The density of iron clusters also reduces with distance until virtually no coalescence occurs. Therefore, particle growth is limited by iron atoms and clusters expansion in space.

Next, we try to describe phenomenologically the process of iron aerosol growth in the hydrocarbon atmosphere. Simultaneously with iron cluster growth, products of $CH_4$ dissociation in the hot arc adhere to the cluster surface releasing carbon atoms. The metal cluster growth is terminated once its surface is fully covered with carbon atoms, i.e., when $\theta = 1$. As experimentally observed, at the chosen $CH_4$ partial pressure and arc parameters used in this work, the carbon shell formation becomes the limiting factor of iron particle growth rather than atoms and clusters expansion in space.

The described above growth saturation can be explained by analyzing Eqs. 1 and 2. When looking into the second term of Eq. 1, a partially carbon-covered iron cluster, $0 < 1-\theta < 1$, would have a limited surface exposed to metal condensation, decreasing the likelihood of iron atom adsorption, and slowing down cluster growth. When fully encapsulated in carbon, i.e., $1-\theta = 0$, adsorption stops, which is represented by the second term of Eq. 1 canceling out. A similar line of thought applies to Eq. 2, a partial coverage by carbon reduces the probability of clusters coalescing with each other, while a fully formed carbon shell, $1-\theta = 0$, cancels out the coalescence term in Eq. 2. Additional carbon atoms are still able to adhere to the shell increasing its thickness, but the iron core growth is ceased.

Apart from the straightforward applicability to core-shell particles, we argue that the process of slowing down the iron particle growth while the shell develops is also relevant to CNT synthesis. In situ TEM observations by Lin et al. showed that the growth rate of a SWCNT on a Ni particle initially increased and then decreased until the growth terminated [52]. They attributed the decreasing growth rate to the continuous coverage of the catalyst with carbon, which passivated the active catalyst sites and gradually reduced its efficiency. In their molecular dynamics study, Ding et al. proposed that more than one carbon island could form on a catalyst surface during the CNT nucleation [53]. Both studies suggest that coverage of the catalyst surface is a dynamic process occurring during tube formation and growth, which, as shown in our model, would inhibit catalyst size growth. We hope that this finding, explained here with a simplified model, will motivate the development of more comprehensive models capable of predicting the kinetics of metal particle growth in hydrocarbon arcs.

## 6 Conclusions

We explored iron nanoparticles growth in the Ar/$CH_4$ arc discharge generated from evaporating steel anode. TEM and EDS analyses revealed that the core-shell iron-carbon particles that were formed in the Ar/$CH_4$ mixture had significantly reduced iron core sizes compared to particles grown in pure Ar. The simplified aerosol growth model was able to explain this result by incorporating a hydrocarbon environment and allowing carbon to adsorb onto the iron core. The process of carbon shell formation inhibited iron atoms adsorption and clusters coalescence, eventually leading to the full encapsulation and cessation of iron core growth. This shows the possibility of leveraging the intrinsic formation of the carbon shell on metal particles in hydrocarbon arc discharges, e.g., for the synthesis of core-shell structures and CNT with prescribed properties.

**Acknowledgments**

This work was supported by the U.S. Department of Energy through contract DE-AC02-09CH11466. We thank Hengfei Gu and Bruce Koel (both Princeton University) for insights on material evaluation, the staff of Princeton University Image and Analysis Center for their help with TEM/EDS, and Timothy Bennett, Aleksandr Merzhevsky, and Nirbhav Chopra (all Princeton Plasma Physics Laboratory) for technical assistance and fruitful discussions.

**References**

[1] K.L. Kelly, E. Coronado, L.L. Zhao, G.C. Schatz, The Optical Properties of Metal Nanoparticles: The Influence of Size, Shape, and Dielectric Environment, J. Phys. Chem. B 107 (2003) 668–677. DOI: 10.1021/jp026731y.

[2] L. Liu, A. Corma, Metal Catalysts for Heterogeneous Catalysis: From Single Atoms to Nanoclusters and Nanoparticles, Chem. Rev. 118 (2018) 4981–5079. DOI: 10.1021/acs.chemrev.7b00776.

[3] B.R. Cuenya, Synthesis and catalytic properties of metal nanoparticles: Size, shape, support, composition, and oxidation state effects, Thin Solid Films 518 (2010) 3127–3150. DOI: 10.1016/j.tsf.2010.01.018.

[4] Y. Xu, Y. Ma, Y. Liu, S. Feng, D. He, P. Haghi-Ashtiani, A. Dichiara, L. Zimmer, J. Bai, Evolution of Nanoparticles in the Gas Phase during the Floating Chemical Vapor Deposition Synthesis of Carbon Nanotubes, The Journal of Physical Chemistry C 122 (2018) 6437–6446. DOI: 10.1021/acs.jpcc.8b00451.

[5] C. Schünemann, F. Schäffel, A. Bachmatiuk, U. Queitsch, M. Sparing, B. Rellinghaus, K. Lafdi, L. Schultz, B. Büchner, M.H. Rümmeli, Catalyst Poisoning by Amorphous Carbon during Carbon Nanotube Growth: Fact or Fiction?, ACS Nano 5 (2011) 8928–8934. DOI: 10.1021/nn2031066.

[6] A. Lu, E.L. Salabas, F. Schüth, Magnetic Nanoparticles: Synthesis, Protection, Functionalization, and Application, Angew Chem Int Ed 46 (2007) 1222–1244. DOI: 10.1002/anie.200602866.

[7] S Odenbach, Recent progress in magnetic fluid research, Journal of Physics: Condensed Matter 16 (2004) R1135. DOI: 10.1088/0953-8984/16/32/R02.

[8] C. Wang, P. Zhai, Z. Zhang, Y. Zhou, J. Ju, Z. Shi, D. Ma, R.P.S. Han, F. Huang, Synthesis of Highly Stable Graphene-Encapsulated Iron Nanoparticles for Catalytic Syngas Conversion, Particle & Particle Systems Characterization 32 (2015) 29–34. DOI: 10.1002/ppsc.201400039.

[9] S. Jing, J. Lu, G. Yu, S. Yin, L. Luo, Z. Zhang, Y. Ma, W. Chen, P.K. Shen, Carbon-Encapsulated WOx Hybrids as Efficient Catalysts for Hydrogen Evolution, Advanced Materials 30 (2018) 1705979. DOI: 10.1002/adma.201705979.

[10] Y. Xu, Y. Shan, Y. Zhang, B. Yu, Y. Shen, H. Cong, Multifunctional Fe3O4@C-based nanoparticles coupling optical/MRI imaging and pH/photothermal controllable drug release as efficient anti-cancer drug delivery platforms, Nanotechnology 30 (2019) 425102. DOI: 10.1088/1361-6528/ab2e40.

[11] M. Nagatsu, T. Yoshida, M. Mesko, A. Ogino, T. Matsuda, T. Tanaka, H. Tatsuoka, K. Murakami, Narrow multi-walled carbon nanotubes produced by chemical vapor deposition using graphene layer encapsulated catalytic metal particles, Carbon 44 (2006) 3336–3341. DOI: 10.1016/j.carbon.2006.06.005.

[12] J.H.J. Scott, S.A. Majetich, Morphology, structure, and growth of nanoparticles produced in a carbon arc, Phys. Rev. B 52 (1995) 12564–12571. DOI: 10.1103/PhysRevB.52.12564.

[13] P.-Z. Si, Z.-D. Zhang, D.-Y. Geng, C.-Y. You, X.-G. Zhao, W.-S. Zhang, Synthesis and characteristics of carbon-coated iron and nickel nanocapsules produced by arc discharge in ethanol vapor, Carbon 41 (2003) 247–251. DOI: 10.1016/S0008-6223(02)00280-4.

[14] N. Aguiló-Aguayo, M.J. Inestrosa-Izurieta, J. García-Céspedes, E. Bertran, Morphological and Magnetic Properties of Superparamagnetic Carbon-Coated Fe Nanoparticles Produced by Arc Discharge, J. Nanosci. Nanotech. 10 (2010) 2646–2649. DOI: 10.1166/jnn.2010.1420.

[15] J. Borysiuk, A. Grabias, J. Szczytko, M. Bystrzejewski, A. Twardowski, H. Lange, Structure and magnetic properties of carbon encapsulated Fe nanoparticles obtained by arc plasma and combustion synthesis, Carbon 46 (2008) 1693–1701. DOI: 10.1016/j.carbon.2008.07.011.

[16] C. Corbella, S. Portal, M.N. Kundrapu, M. Keidar, Nanosynthesis by atmospheric arc discharges excited with pulsed-DC power: a review, Nanotechnology 33 (2022) 342001. DOI: 10.1088/1361-6528/ac6bad.

[17] A.V. Eremin, E.V. Gurentsov, S.A. Musikhin, Synthesis of binary iron–carbon nanoparticles by UV laser photolysis of $Fe(CO)_5$ with various hydrocarbons, Materials Research Express 3 (2016) 105041. DOI: 10.1088/2053-1591/3/10/105041.

[18] J.B. Park, S.H. Jeong, M.S. Jeong, J.Y. Kim, B.K. Cho, Synthesis of carbon-encapsulated magnetic nanoparticles by pulsed laser irradiation of solution, Carbon 46 (2008) 1369–1377. DOI: 10.1016/j.carbon.2008.05.011.

[19] A.A. El-Gendy, E.M.M. Ibrahim, V.O. Khavrus, Y. Krupskaya, S. Hampel, A. Leonhardt, B. Büchner, R. Klingeler, The synthesis of carbon coated Fe, Co and Ni nanoparticles and an examination of their magnetic properties, Carbon 47 (2009) 2821–2828. DOI: 10.1016/j.carbon.2009.06.025.

[20] N. Luo, K. Liu, Z. Liu, X. Li, S. Chen, Y. Shen, T. Chen, Controllable synthesis of carbon coated iron-based composite nanoparticles, Nanotechnology 23 (2012) 475603. DOI: 10.1088/0957-4484/23/47/475603.

[21] J. Zheng, Z.Q. Liu, X.S. Zhao, M. Liu, X. Liu, W. Chu, One-step solvothermal synthesis of $Fe_3O_4$@C core–shell nanoparticles with tunable sizes, Nanotechnology 23 (2012) 165601. DOI: 10.1088/0957-4484/23/16/165601.

[22] B. Liu, Y. Shao, X. Xiang, F. Zhang, S. Yan, W. Li, Highly efficient one-step synthesis of carbon encapsulated nanocrystals by the oxidation of metal π-complexes, Nanotechnology 28 (2017) 325603. DOI: 10.1088/1361-6528/aa7995.

[23] Z. Shi, Y. Lian, F.H. Liao, X. Zhou, Z. Gu, Y. Zhang, S. Iijima, H. Li, K.T. Yue, S.-L. Zhang, Large scale synthesis of single-wall carbon nanotubes by arc-discharge method, Journal of Physics and Chemistry of Solids 61 (2000) 1031–1036. DOI: 10.1016/S0022-3697(99)00358-3.

[24] C. Journet, W.K. Maser, P. Bernier, A. Loiseau, M.L. De La Chapelle, S. Lefrant, P. Deniard, R. Lee, J.E. Fischer, Large-scale production of single-walled carbon nanotubes by the electric-arc technique, Nature 388 (1997) 756–758. DOI: 10.1038/41972.

[25] S. Iijima, Helical microtubules of graphitic carbon, Nature 354 (1991) 56–58. DOI: 10.1038/354056a0.

[26] S. Yatom, R.S. Selinsky, B.E. Koel, Y. Raitses, "Synthesis-on" and "synthesis-off" modes of carbon arc operation during synthesis of carbon nanotubes, Carbon 125 (2017) 336–343. DOI: 10.1016/j.carbon.2017.09.034.

[27] M Keidar, A M Waas, On the conditions of carbon nanotube growth in the arc discharge, Nanotechnology 15 (2004) 1571. DOI: 10.1088/0957-4484/15/11/034.

[28] C. Journet, M. Picher, V. Jourdain, Carbon nanotube synthesis: from large-scale production to atom-by-atom growth, Nanotechnology 23 (2012) 142001. DOI: 10.1088/0957-4484/23/14/142001.

[29] C.L. Cheung, A. Kurtz, H. Park, C.M. Lieber, Diameter-Controlled Synthesis of Carbon Nanotubes, J. Phys. Chem. B 106 (2002) 2429–2433. DOI: 10.1021/jp0142278.

[30] A.G. Nasibulin, P.V. Pikhitsa, H. Jiang, E.I. Kauppinen, Correlation between catalyst particle and single-walled carbon nanotube diameters, Carbon 43 (2005) 2251–2257. DOI: 10.1016/j.carbon.2005.03.048.

[31] X. Zhang, B. Graves, M. De Volder, W. Yang, T. Johnson, B. Wen, W. Su, R. Nishida, S. Xie, A. Boies, High-precision solid catalysts for investigation of carbon nanotube synthesis and structure, Sci. Adv. 6 (2020) eabb6010. DOI: 10.1126/sciadv.abb6010.

[32] G. Chen, Y. Seki, H. Kimura, S. Sakurai, M. Yumura, K. Hata, D.N. Futaba, Diameter control of single-walled carbon nanotube forests from 1.3–3.0 nm by arc plasma deposition, Scientific Reports 4 (2014) 3804. DOI: 10.1038/srep03804.

[33] P.L. Johnson, R.J. Hanson, S. Carlos, R.W. Taylor, PLASMA REACTOR. US patent 9,574,086 B2, 2017.

[34] Predtechenskiy, METHOD AND APPARATUS FOR PRODUCING CARBON NANOSTRUCTURES. US patent 2020/0239316 A1, 2020.

[35] X.L. Dong, Z.D. Zhang, Q.F. Xiao, X.G. Zhao, Y.C. Chuang, S.R. Jin, W.M. Sun, Z.J. Li, Z.X. Zheng, X.L. Dong, Characterization of ultrafine -Fe(C), -Fe(C) and Fe3C particles synthesized by arc-discharge in methane, Journal of Materials Science 33 (1998) 1915–1919. DOI: 10.1023/A:1004369708540.

[36] C. Hao, F. Xiao, Z. Cui, Preparation and structure of carbon encapsulated copper nanoparticles, Journal of Nanoparticle Research 10 (2008) 47–51. DOI: 10.1007/s11051-007-9218-6.

[37] X. Zhang, Y. Rao, J. Guo, G. Qin, Multiple-phase carbon-coated FeSn2/Sn nanocomposites for high-frequency microwave absorption, Carbon 96 (2016) 972–979. DOI: 10.1016/j.carbon.2015.09.087.

[38] V. Sunny, D. Sakthi Kumar, Y. Yoshida, M. Makarewicz, W. Tabiś, M.R. Anantharaman, Synthesis and properties of highly stable nickel/carbon core/shell nanostructures, Carbon 48 (2010) 1643–1651. DOI: 10.1016/j.carbon.2010.01.006.

[39] N.C. Roy, M.R. Talukder, Effect of pressure on the properties and species production in gliding arc Ar, O2, and air discharge plasmas, Physics of Plasmas 25 (2018) 093502. DOI: 10.1063/1.5043182.
[40] B. Syed, T.-W. Hsu, A.B.B. Chaar, P. Polcik, S. Kolozsvari, G. Håkansson, J. Rosen, L.J.S. Johnson, I. Zhirkov, J.M. Andersson, M.J. Jöesaar, M. Odén, Effect of varying N2 pressure on DC arc plasma properties and microstructure of TiAlN coatings, Plasma Sources Science and Technology 29 (2020) 095015. DOI: 10.1088/1361-6595/abaeb4.
[41] M. Tanaka, S. Tashiro, T. Satoh, A.B. Murphy, J.J. Lowke, Influence of shielding gas composition on arc properties in TIG welding, Science and Technology of Welding and Joining 13 (2008) 225–231. DOI: 10.1179/174329308X283929.
[42] S. Gershman, Y. Raitses, Unstable behavior of anodic arc discharge for synthesis of nanomaterials, J. Phys. D: Appl. Phys. 49 (2016) 345201. DOI: 10.1088/0022-3727/49/34/345201.
[43] F. Liang, M. Tanaka, S. Choi, T. Watanabe, Formation of different arc-anode attachment modes and their effect on temperature fluctuation for carbon nanomaterial production in DC arc discharge, Carbon 117 (2017) 100–111. DOI: 10.1016/j.carbon.2017.02.084.
[44] V. Vekselman, M. Feurer, T. Huang, B. Stratton, Y. Raitses, Complex structure of the carbon arc discharge for synthesis of nanotubes, Plasma Sources Sci. Technol. 26 (2017) 065019. DOI: 10.1088/1361-6595/aa7158.
[45] A. Khrabry, I.D. Kaganovich, A. Khodak, V. Vekselman, T. Huang, Analytical model of low and high ablation regimes in carbon arcs, Journal of Applied Physics 128 (2020) 123303. DOI: 10.1063/5.0016110.
[46] A. Gutsch, H. Mühlenweg, M. Krämer, Tailor-Made Nanoparticles via Gas-Phase Synthesis, Small 1 (2004) 30–46. DOI: 10.1002/smll.200400021.
[47] C.M. Wang, D.R. Baer, L.E. Thomas, J.E. Amonette, J. Antony, Y. Qiang, G. Duscher, Void formation during early stages of passivation: Initial oxidation of iron nanoparticles at room temperature, Journal of Applied Physics 98 (2005) 094308. DOI: 10.1063/1.2130890.
[48] M.C. Diaz, H. Jiang, E. Kauppinen, R. Sharma, P.B. Balbuena, Can Single-Walled Carbon Nanotube Diameter Be Defined by Catalyst Particle Diameter?, J. Phys. Chem. C 123 (2019) 30305–30317. DOI: 10.1021/acs.jpcc.9b07724.
[49] N. Arora, N.N. Sharma, Arc discharge synthesis of carbon nanotubes: Comprehensive review, Diamond and Related Materials 50 (2014) 135–150. DOI: 10.1016/j.diamond.2014.10.001.
[50] V.A. Nemchinsky, M. Shigeta, Simple equations to describe aerosol growth, Modelling and Simulation in Materials Science and Engineering 20 (2012) 045017. DOI: 10.1088/0965-0393/20/4/045017.
[51] S.L. Girshick, C. Chiu, Kinetic nucleation theory: A new expression for the rate of homogeneous nucleation from an ideal supersaturated vapor, The Journal of Chemical Physics 93 (1990) 1273–1277. DOI: 10.1063/1.459191.
[52] M. Lin, J.P. Ying Tan, C. Boothroyd, K.P. Loh, E.S. Tok, Y.-L. Foo, Direct Observation of Single-Walled Carbon Nanotube Growth at the Atomistic Scale, Nano Lett. 6 (2006) 449–452. DOI: 10.1021/nl052356k.

[53] F. Ding, K. Bolton, A. Rosén, Nucleation and Growth of Single-Walled Carbon Nanotubes: A Molecular Dynamics Study, J. Phys. Chem. B 108 (2004) 17369–17377. DOI: 10.1021/jp046645t.